\documentclass{article}
\usepackage{amssymb, amsmath, latexsym}
\usepackage[dvips]{epsfig}
\newtheorem{theorem}{Theorem}
\newtheorem{lemma}{Lemma}
\newtheorem{proposition}{Proposition}

\begin{document}
\title{\bf Parallel Chip Firing Game associated with $n$-{\em cube} orientations}  
\author{Ren\'e Ndoundam${}^{1}$, Maurice Tchuente${}^{1,2}$, Claude Tadonki${}^{3}$ \\
${}^{1}$ {\small Department of Computer Science, Faculty of Science, 
 University of Yaound\'e I,  } \\
{\small P.o. Box. 812 Yaound\'e, Cameroon} \\
${}^{2}$ {\small Department of Mathematics and Computer Science, Faculty of Science, University of Ngaound\'er\'e } \\
${}^{3}${\small University of Paris-Sud 11 ,  CNRS , IEF (AXIS Group) } \\
        {\small UMR 8622 - B\^at 220 - Centre Scientifique d'Orsay - F91405 Orsay Cedex , France}    \\
{\small E.mail : ndoundam@yahoo.com , Maurice.Tchuente@ens-lyon.fr , claude.tadonki@u-psud.fr } \\
                 } 
\maketitle
\begin{abstract} We study the cycles generated by the chip firing game
 associated with $n$-{\em cube} orientations. We show the existence of the cycles generated by
{\em parallel evolutions} of even lengths from $2$ to $2^n$ on $H_n$ ($n\geq 1$), and of odd lengths different
 from $3$ and ranging from $1$ to $2^{n{\rm -}1}{\rm -}1$ on $H_n$ ($n\geq 4$).
\end{abstract}

\pagestyle{myheadings}
\thispagestyle{plain}
\markboth{R. Ndoundam, M. Tchuente and C. Tadonki}{Parallel Chip firing game
  on hypercube}

\section{Introduction}
Consider a digraph $G=(V,A)$, where $V=\{ 1,...,n\}$ is the set of vertices
and $A\subseteq V\times V$ is the set of arcs. The {\em out-degree} ( resp. {\em in-degree} ) of a
vertex $i$, hereafter denoted by $d^+(i)$ (resp. $d^-(i)$ ), is the number of vertices $j$ such
that $(i,j)\in A$ (resp. $(j,i)\in A$). A vertex
with {\em out-degree} zero is called a {\em sink}. All these notions apply to an
undirected graph $G=(V,E)$ by considering an edge $e=[i,j]$ as two opposite
arcs $(i,j)$ and $(j,i)$.

In the {\em parallel chip firing game} played on $G$, a state is a mapping
$x: V \rightarrow N$ which can be viewed as a distribution of chips onto the
vertices of $G$. A vertex is said to be {\em active} in a state $x$ if
$x(i)\geq d^+(i)$, otherwise it is said to be {\em passive}. In a move of the
game, a state $x$ is transformed into a new state as follows : every vertex
tries to send one chip to every {\em out-neighbor}.

\hspace{2mm}$\bullet$ If it is not possible, i.e. if $i$ is {\em passive}, then it
resigns ;

\hspace{2mm}$\bullet$ Otherwise, vertex $i$ is {\em active} and sends the chips.\\
It is easily seen that the number of chips remains constant. Therefore, the
evolution is ultimately {\em periodic}. More precisely, if $x^t,t\geq 0$,
denotes the state of the system at time $t$, then there exists an integer $q$
called {\em transient length} and another integer $p$ called {\em period}
or  {\em cycle length } such that \\
\begin{equation}  \label{eqn:period}
x^{t+p}=x^t \mbox{ for } t\geq q, \mbox{ and } x^{t+p'}\neq x^t \mbox{ for }
p'<p.
\end{equation}\\
The sequence $x^0,x^1,...,x^{q-1}$ is called the {\em transient} and every
sequence of $p$ consecutive states $x^t,x^{t+1},...,x^{t+p-1}$, such that
$t\geq q$, is called a {\em cycle} of the evolution.

Following Spencer's introductory paper \cite{SPE:86} which was devoted to the
chip firing game on chains, many authors have been interested in this
problem. The most interesting questions concern the relationships between the
structure of the graph on one hand, and the transients and periods generated
by the chip firing game on the other hand. Concerning the {\em period}, Bitar and
Goles have shown that if $G$ is a tree, then only periods one and two occur
\cite{BIT:92}. Later, Prisner has studied a generalization of the game by
considering multigraphs, i.e. digraphs with multiplicities on the arcs. He
has then shown that there is a sharp contrast in the behavior for eulerian
digraphs (i.e. digraphs where the {\em in-degre} of each vertex equals it {\em
  out-degree}). More precisely, he has proved that in every strongly connected
euleurian multigraph, any divisor of every dicycle length occurs as a period
\cite{PRI:94}. He has also shown that there is no polynomial $h(n)$ such that
the periods generated by the chip firing game on digraphs of order $n$ are
bounded by $h(n)$. Readers interested by other results on periods and
transients of the chip firing game may refer to \cite{TAR:88,AND:89,BIT:89,BJO:91,ERI:91,GOL:93}.
 Readers interested by combinatorial games may refer to \cite{Gol:02, GM:02, Gol:04, Sjo:05, Fra:09}.

There is a particular case where the chip firing game is related to graph
orientations. Indeed, let us consider an undirected graph $G=(V,E)$, and let
us
assume that initially, the edges of $G$ can be oriented in such a way that the
number of chips of every vertex equals the {\em in-degree} of that vertex. If
this property is true in the initial configuration, then it remains true throughout the
game. One step of the game then consists in reversing the orientations of all
edges going into sinks. Goles and Prisner \cite{GOL:00} have studied {\em gardens
of Eden}, i.e. states that can appear only at time $t=0$. They have also
studied the relationships between graph orientations and evolutions induced by
states with $|E|$ chips. Moreover, Kiwi, Ndoundam, Tchuente and Goles
\cite{KIW:94} have exhibited cycles of exponential length $e^{\Omega (n\log n)}$
generated by the chip firing game associated with the orientations of cascades
of rings. Other results on this particular case may be found in \cite{ERI:94}.

In this paper, we study the dynamics generated by the chip firing game
associated with $n$-$cube$ orientations. More precisely, using a
recurrent approach, we show that for $n\geq 4$, there exists cycles of even lengths from
 $2$ to $2^n$ on $H_n$ ($n\geq 1$), and of odd lengths different from $3$ and ranging from $1$ to
 $2^{n{\rm -}1}{\rm -}1$ on $H_n$ ($n\geq 4$).

The remainder of this paper is organized as follows. In the next section, we
present some basic notations and definitions related to $n$-$cubes$. Section 3 is devoted to the recurrent
 construction of {\em left cyclic partitions} and possible period lengths whereas section 4 presents some concluding remarks.

\section{Basic notations and definitions}
An {\em n-dimensional hypercube} (or $n${\em -cube}) is an undirected graph
$H_n=(V,E)$, where $V=\{ 0,1 \} ^n$ is the set of vertices and two nodes
$u=(u_0,u_1,...,u_{n-1})$ and $v=(v_0,v_1,...,v_{n-1})$ are neighbors if and
only if they differ in only one bit in their binary representations, i.e.
there is an integer $i$ such that $u_i\neq v_i$ and $u_j=v_j$ for $j\neq
i$. One can define recursively the $n${\em -cube} as follows :

$\bullet$ The $0$-{\em cube} is reduced to one vertex ;

$\bullet$ $H_{n+1}$ is obtained by taking two copies of $H_{n}$ and connecting
all pairs of equivalent vertices.\\
Fig. 1 illustrates this constructions for $n=0,1,2,3$.
\begin{figure}[htbp]
\centering
\epsfxsize=8.6cm
\epsfbox{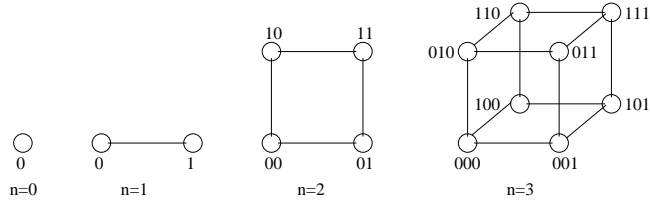}
\caption{\label{figgraph} $n$-{\em cubes} for $n\leq 3$.}
\end{figure}

Hereafter, given a set $W$ and a boolean value $x$, we denote $xW=\{ xu:u\in
W\}$. With this notation, we can write $H_{n+1}=0H_{n}\cup 1H_n$.

{\bf Definition 1.} A {\em block-sequential evolution} of the chip firing game
associated with graph orientations and played on an $n$-{\em cube} is
obtained as follows. Consider a sequence of non empty subsets $\{ W_i;i\geq
0\}$ of $\{ 0,1\} ^n$. At time $t$, every vertex $u$ of $W_t$ is
considered. If $u$ is a sink then the orientation of all its {\em in-going}
arcs are reversed, otherwise no action is undertaken. Hereafter, we say that a vertex {\em fires} at time $t$
if it belongs to $W_t$ and is a {\em sink} at time $t$.\vspace{2mm} \\

The {\em parallel evolution} is therefore a particular case of the general
scheme described above. Another classical evolution scheme is the {\em
  sequential evolution} where $W_t$ is reduced to one vertex (i.e. $|W_t|=1$)
and there is a permutation $\sigma$ of $\{ 0,...,2^n{\rm -}1 \}$ such that $\{
W_i;i\geq 0\}$ is periodic of period $W_{\sigma(0)}, W_{\sigma(1)},...,
W_{\sigma(2^n{\rm-}1)}$. Both parallel and sequential evolutions are particular cases of the
so-called {\em serial-parallel} evolutions where the sequence $\{
W_i;i\geq 0 \}$ is periodic of period $W_0, W_1,...,W_{k-1}$, with the constraint that
  $W_0\cup W_1\cup ... \cup W_{k{\rm -}1}$ is a partition of $\{ 0,1,...,2^n{\rm
    -}1 \}$.\vspace{2mm}\\

{\bf Definition 2.} A partition $S_0\cup S_1\cup ...\cup S_{k-1}$ of the
vertices of an $n$-{\em cube} is called a {\em left cyclic} partition if the
two following statements hold.

$\bullet$ For all $i$ from $0$ to $k{\rm -}1$, every vertex of $S_i$ has a
neighbor in $S_{i-1}$, where index operations are performed modulo $k$.

$\bullet$ For all $i$ from $0$ to $k{\rm -}1$, there is no edge between two
vertices of $S_i$.\vspace{2mm}\\
{\bf Comment 2.} Canonical decompositions defined in \cite{GOL:00} for
acyclic digraphs are obtained from {\em left cyclic partitions} by orienting
the edges such that all arcs from the set $S_i$ go to sets
$S_j$ such that $j>i$. On the other hand, {\em left cyclic partitions} are more
restrictive than the partitions introduced in \cite{PRI:94} since we do not
allow edges joining two vertices of the same subset. Indeed, in the chip
firing game associated with graph orientations, two neighbors cannot fire
simultaneously, whereas this situation is possible for the general chip firing
game. \\

We present an important property of left cyclic partitions on an $n$-{\em cube}.

\begin{theorem}
If a partition $S_0\cup S_1\cup ...\cup S_{k-1}$ of the vertices of an $n$-
  cube $H_n$ is a left cyclic partition then there is a
  cyclic evolution of the chip firing game associated with graph orientations
  and played on $H_n$, such that for every $t\geq 0$, $S_t$ is the set of
  vertices which are fired at time $t$.
\end{theorem}
{\bf Proof.} Let $S_0,S_1,...,S_{k-1}$ be a {\em left cyclic partition}. Consider
an orientation where every edge $e=[u,v]$ such that $u\in S_i$ and $v\in
S_j,i<j$, is oriented from $v$ to $u$. It is easily seen that in the parallel chip
firing game starting with such a configuration, the subsets of vertices
which fire at successive steps correspond to a periodic sequence of period
$S_0,S_1,...,S_{k-1}$.\\
$\Box$

\section{Recurrent construction of left cyclic partitions}
In this section, we first present the construction of {\em left cyclic
  partitions} of even lengths.
\begin{lemma}   \label{lem:allpair}
An $n$-cube admits left cyclic partitions of all even lengths from
$2$ to $2^n$.
\end{lemma}
{\bf Proof.} Let $H_n=(V,E)$ be an $n$-{\em cube} an let $p$ be an even
integer between $2$ and $2^n$. It is well known that, since $p$ is even,
there is a cycle $[x_0,x_1, ...,x_{p-1}, x_0]$
of length $p$ in $H_n$. Now, for every vertex $u$, let $\Gamma (u)$ denote the
set of all neighbors of $u$ in $H_n$. This notation is naturally extended to a
set of vertices. A {\em left cyclic partition} of order $p$ is
obtained  as follows.

\ {\bf For}  $i=0,...,p-1$ {\bf do}

\ \ \ $S_i\leftarrow \{ x_i\}$

\ {\bf endfor}

\ $S=V-\{ x_0,x_1,...x_{p-1} \}$

\ {\bf while} $(S\neq \emptyset)$ {\bf do}

\ \ \ {\bf For } $i\leftarrow 0$ {\bf to } $p-1$  {\bf do}

\ \ \ \ \hspace{2mm} $S_{i+1}\leftarrow S_{i+1}\cup (\Gamma (S_i)\cap S)$

\ \ \ \ \hspace{2mm} $S\leftarrow S-(\Gamma (S_i)\cap S)$

\ \ \ \ {\bf endfor}

\ {\bf endwhile}

It is obvious that $S_0,...,S_{p-1}$ is a partition of $V$ and that every
vertex in $S_i$ has at least one neighbor in $S_{i-1}$. So we just need to show
that two vertices of the same subset $S_i$ cannot be neighbors. Let $a$ and
$b$ be two vertices of $S_i$.

$\bullet$ There is a path from a to $x_0$ of length $\ell _1$ such that $\ell
_1=i$ {\bf  mod }$p$,

$\bullet$ There is a path from b to $x_0$ of length $\ell _2$ such that $\ell
_2=i$ {\bf  mod }$p$,\\
Since $p$ is  even, it follows that $\ell _1=\ell _2$ {\bf mod }$2$. Hence, if
$a$ and $b$ were neighbors, there would exist a cyclic path of odd length
${\ell}_1+{\ell}_2+1$ joining $a$ and $b$ in $H_n$, which is not possible
since $H_n$ is a bipartite graph. This shows that two vertices of the same
subset cannot be neighbors.\\
$\Box$.\\
The following figure displays the partition of order $4$ in $H_3$ obtained by
the previous procedure starting with the cycle $[000,001,011,010]$.
\begin{figure}[htbp]
\centering
\epsfxsize=6cm
\epsfbox{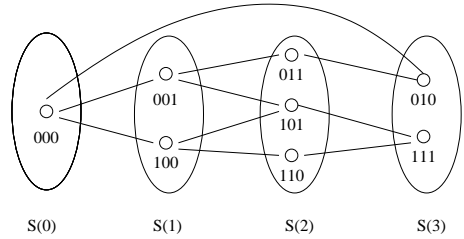}
\caption{\label{tracegraph} {\em Left cyclic partition} of orders $4$ generated in $H_3$.}
\end{figure}\\
Let us now turn to the construction of {\em left cyclic partitions} of odd
lengths.

\begin{lemma}  \label{lem:23}
If $S_0,S_1,S_2$ is a {\em left cyclic partition} of $H_n,n\geq 2$, then every vertex
of $S_i$ has at least two neighbors in $S_{i-1}$ for $i=0,1,2$.
\end{lemma}
{\bf Proof.} Because of symmetry considerations, we can assume that $i=2$. So
let $x$ be a vertex of $S_2$. From the definition of {\em left cyclic
  partitions},

$\bullet$ $x$
has a neighbor $x\oplus e_j$ in $S_1$, where $\oplus$ is the {\sc xor}
operator and $e_j$ is a

\ \ vector of the canonical basis.

$\bullet$ similarly, $x\oplus e_j$  has a neighbor
$x\oplus e_j\oplus e_k$ in $S_0$. \\
Now consider the vertex $x\oplus e_k$. 

$\bullet$ It is a neighbor of $x$, hence it does not belong to $S_2$.

$\bullet$ It is a neighbor of $x\oplus e_j\oplus e_k$, hence it does not belong to $S_0$.\\
It then follows that $x\oplus e_k$ belongs to $S_1$, hence $x$ admits two
neighbors $x\oplus e_j$ and $x\oplus e_k$ which are both in $S_1$.\\
$\Box$\\

\begin{lemma} \label{lem:3rec}
If $H_n$, $n\geq 3$ admits a {\em left cyclic partition} of order $3$, then 
$H_{n-1}$ admits a {\em left cyclic partition} of order $3$.
\end{lemma}
{\bf Proof.} Let $S_0,S_1,S_2$ be a {\em left cyclic partition} of order $3$ of 
$H_n$. Let $x$ be a vertex of $S_i$. We can assume without loss of generality
that $x=1a$. Since $H_n=0H_{n-1}\cup 1H_{n-1}$, 
$y=0a$ is the unique neighbor of $x$ in $0H_{n-1}$. Consequently, from lemma \ref{lem:23}, $x$ admits a neighbor in $1H_{n-1} \cap S_{i-1}$. This shows that the
subgraph $1H_{n-1}$ which is isomorphic to $H_{n-1}$, contains a {\em left
  cyclic partition} of order $3$.\\
$\Box$

\begin{proposition}  \label{prop:pas3}
$n$-cubes do not  admit left cyclic partitions of order 3. 
\end{proposition}
{\bf Proof.} An $n$-{\em cube} with $n\leq 1$ has less than $3$ vertices and
 cannot admit a {\em left cyclic partition} of order $3$. On the other hand,
 from lemma \ref{lem:23}, if $S_0,S_1,S_2$ is a {\em left cyclic partition} of
 an $n$-{\em cube}, $n\geq 2$, then every $S_i$ contains at least two elements
 (i.e. $|S_i|\geq 2$). Consequently, the $2$-{\em cube} $H_2$ which is of
 cardinality $4$ cannot admit a {\em left cyclic partition} of order $3$. By application of lemma
 \ref{lem:3rec}, we deduce that no $n$-{\em cube}, $n\geq 3$ admits a {\em left
 cyclic partition} of order $3$.\\
 $\Box$
\\
Proposition \ref{prop:pas3} gives the lower bound for {\em left cyclic partitions} of odd lengths. 
Let now study the upper bound.
\begin{proposition}   \label{prop:ibound}
If $S_0,...,S_{p-1}$ is a left cyclic partition of odd order $p$ of $H_n$, then 
$p\leq 2^{n-1}-1$. 
\end{proposition}
{\bf Proof.} We just have to show that in such a case,   
$|S_i|\geq 2$ for $i=0,...,p-1$. Indeed, starting from a vertex $a_{p-1}\in S_{p-1}$, 
we construct a chain $[a_{p-1}$,$ a_{p-2}$,$ ...$ , 
$a_0$, $b_{p-1}$, $b_{p-2}$, $...$, $b_0]$ such 
that $a_i,b_i\in S_i$ for $i=0,...,p{\rm -}1$. It is clear that $a_i\neq b_i,i=0,...,p-1$, otherwise 
we would have displayed a closed path of odd length in $H_n$ which is not
possible.\\
 $\Box$
\\
Now that we have established lower and upper bounds for {\em left cyclic
  partitions} of odd lengths, let us show that 
all intermediate lengths are admissible.
\begin{lemma}   \label{lem:rec}
If $H_n$ admits a {\em left cyclic partition} of order $p$, then $H_{n+1}$ 
admits left cyclic partition of order $p$.
\end{lemma}
{\bf Proof.} If $S_0,...,S_{p-1}$ is a left cyclic partition of order 
$p$ in $H_n$, then it is easily checked that $1S_i\cup 0S_{i-1},i=0,...,p-1$ is
a {\em left cyclic 
partition} of order $p$ in $H_{n+1}$\\
 $\Box$.

\begin{lemma}   \label{lem:suiv}
If $H_n$ admits a {\em left cyclic partition} of odd order $p$, $p\geq 5$ then $H_{n+1}$  
admits a {\em left cyclic partition} of order $2p-1$. Moreover, if $p\geq 7$, then 
$H_{n+1}$ admits a left cyclic partition of order $2p-3$.
\end{lemma}
{\bf Proof.} Let $S_0,S_1,...,S_{p-1}$ be a {\em left cyclic partition} of
odd order $p$.

$\bullet$ Case $p\geq 5$\\
The following sequence is a {\em left cyclic partition} of order $2p{\rm -}1$ in
$H_{n+1}$.\\
$0S_0$, $1S_0\cup 0S_1$, $1S_1$, $1S_2$, $0S_2$, $0S_3$, $1S_3$
$...$,
 $1S_{2i}$, $0S_{2i}$, $0S_{2i+1}$, $1S_{2i+1}$,..., $1S_{p-3}$, $0S_{p-3}$,
 $0S_{p-2}$, $1S_{p-2}$, $1S_{p-1}$, $0S_{p-1}$,\\ 

$\bullet$ Case $p\geq 7$\\
A {\em left cyclic partition} of order $2p{\rm -}3$ in
$H_{n+1}$ is obtained from the {\em left cyclic partition} exhibited in the
case $p\geq 5$ by replacing the subsequence 
$1S_2$, $0S_2$, $0S_3$, $1S_3$, $1S_4$, $0S_4$, $0S_5$, $1S_5$ 
by 
 $1S_2$, $0S_2\cup 1S_3$, $0S_3$, $0S_4$, $1S_4\cup 0S_5$, $1S_5$.\\
$\Box$

\begin{lemma}   \label{lem:57}
$H_4$ admits left 
cyclic partitions of orders $5$ and $7$.
\end{lemma}
{\bf Proof.}\\
$\bullet$  A {\em left cyclic partition} of order $5$ in $H_4$ is the following
:\\
$\{ 0000,1101\}$, $\{ 0001,1100,0010,1111\}$, 
$\{ 0110,1011\}$, $\{ 0100,0111,1001,1010\}$,\\ 
$\{ 0011,0101,1000,1110\}$.\\
$\bullet$ A {\em left cyclic partition} of order $7$ in $H_4$ is the following
:\\
$\{ 0000,1101\}$, $\{ 0001,1100\}$, 
$\{ 0011,1110\}$, $\{ 0010,1111\}$, 
$\{ 0110,1011\}$, \\$\{ 0100,0111,1001,1010\}$, 
$\{ 0101,1000\}$.\\
Fig. 3 displays the partitions. 
\begin{figure}[htbp]
\centering
\epsfxsize=8.6cm
\epsfbox{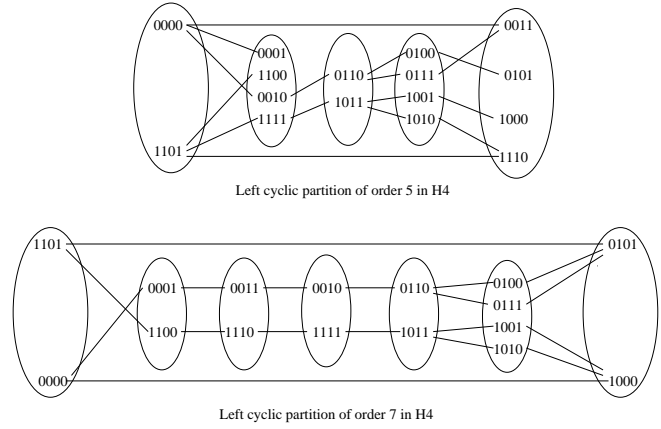} 
\caption{\label{partgraph} {\em Left cyclic partitions} of orders $5$ and $7$ in $H_4$.}
\end{figure}\\
$\Box$

\begin{lemma}  \label{lem:ipmax}
$H_n$, $n\geq 4$, admits a {\em left 
cyclic} partition of order $2^{n-1}-1$.
\end{lemma}
{\bf Proof.} Consider the sequence $\{ u_i;0\leq i\leq 2^{n-1}{\rm -}1\}$,
defined by $u_i=bin(i)\oplus bin(i/2)$, where $bin(x)$ is the
$n$-position binary representation of the integer $x$, and symbol $/$ denotes integer
division. It can be easilly checked that this sequence corresponds to a {\em
  hamiltonian cycle} in $H_{n-1}$. Now, let us denote $v_i=u_i\oplus 1\oplus
2^{n-2}$ (i.e. $v_i$ is obtained from $u_i$ by changing the first and last
bits) and $N=2^n$. It is also easy to check that $\{ v_i;0\leq i\leq
2^{n-1}-1\}$ is a {\em hamiltonian cycle} of $H_{n-1}$. Let us now consider the following
sets\\
\begin{equation}    \label{eqn:bigleft}
\{0u_0,1v_0\},...,
\{0u_{N-4},1v_{N-4}\},\{0u_{N-3},0u_{N-1},1v_{N-3},1v_{N-1}\},
\{0u_{N-2},1v_{N-2}\}. 
\end{equation}\\
At this step, it is important to recall that two vertices referenced by $i$
and $j$ are neighbors in the hypercube if and only if there is an integer $k$
such that $i\oplus j=2^k$. Observe that $0u_i\oplus 1v_i=2^{n-1}\oplus
(u_i\oplus v_i)=2^{n-1}\oplus 1\oplus 2^{n-2}$. Hence, $0u_i$ and $1v_i$ are
not neighbors in the hypercube $H_{n}$ . On the other hand, $u_{N-4}=100...010$,
 $u_{N-2}=10...01$, $u_{N-1}=10...0$ and $v_0=u_0\oplus 1\oplus
 2^{n-2}=10...01=u_{N-2}$. Hence, \\
\begin{equation}   \label{eqn:endc}
0u_{N-4}, 0u_{N-1},  0u_{N-2}, 1v_0
 \mbox{ is a chain of } H_n.
\end{equation}\\
Moreover, $v_{N-4}=0...011$, $v_{N-2}=0...0=u_0$ and
$v_{N-1}=0...01=u_1$. Hence 
\begin{equation}   \label{eqn:firstc}
1v_{N-4}, 1v_{N-1},  1v_{N-2}, 0u_0
 \mbox{ is a chain of } H_n.
\end{equation}\\
Properties \ref{eqn:endc} and \ref{eqn:firstc} together with the fact that
$\{ u_i;0\leq i\leq 2^{n-1}{\rm -}1\}$ and $\{ v_i;0\leq i\leq 2^{n-1}{\rm
  -}1\}$ are both {\em hamiltonian cycles} of $H_n$, imply that the
partition exhibited in \ref{eqn:bigleft} is a {\em left cyclic partition}. \\    
$\Box$ \\
\begin{proposition}   \label{prop:ip}
$H_n$, $n\geq 4$, admits left cyclic partitions of all odd orders 
from $5$ to $2^{n-1}-1$. 
\end{proposition}
{\bf Proof.} We proceed by induction on $n$.
For $n=4$ the result follows from lemma \ref{lem:57}.\\
Assuming that the result holds for $n\geq 4$, let us consider an $(n{\rm
  +}1)$-{\em cube} together with an odd integer $p\in [5,2^n-1]$.

$\bullet$ Case 1 : $5\leq p\leq 2^{n-1}-1$. The result follows from the 
induction hypothesis by application of lemma \ref{lem:rec}.

$\bullet$ Case 2 : $2^{n-1}{\rm -}1<p<2^n-1$. There is an odd integer $q$, 
$7<q< 2^{n-1}-1$, such that $p=2q-1$ or  
$p=2q-3$. The result follows from the induction hypothesis by application of
lemma \ref{lem:suiv}.

$\bullet$ Case 3 : $p=2^n-1$. The result follows from lemma \ref{lem:ipmax}.\\
$\Box$
\\

We are now ready to state the main theorem.

\begin{theorem}  \label{th:main}
There exists cycles generated by the parallel chip firing game associated with n-cube
orientations, $n\geq 4$, are of even lengths from $2$ to $2^n$, and of odd
lengths different from 3 and ranging from $1$ to $2^{n-1}{\rm -}1$.
\end{theorem}
{\bf Proof.} We just need to show that this property holds for {\em left
  cyclic partitions} of vertices of {\em n-cubes}. The existence of {\em left cyclic partitions} of all even lengths
 from $2$ to $2^n$ follows from lemma \ref{lem:allpair}. Let us turn to odd periods $p$. \\
$\bullet$ Case 1 : $p=1$. Consider and orientation which contains a {\em
  hamiltonian cycle}. Clearly, such a configuration is a {\em fixed point} for
  the chip firing game associated with graph orientations. \\
$\bullet$ Case 2 : $p=3$. The non existence of period $3$ follows from proposition \ref{prop:pas3}. \\
$\bullet$ Case 3 : $5\leq p\leq 2^{n-1}-1$. The existence of this period
follows from proposition \ref{prop:ip}. \\
$\Box$

\section{Conclusion}
 We show in the particular case of {\em parallel evolutions} on $n$-{\em cube}, 
 the existence of cycles of even lengths from $2$ to $2^n$, and of odd lengths different
 from $3$ and ranging from $1$ to $-1+2^{n-1}$. In case of {\em parallel evolutions} on 
 $n$-{\em cube}, the existence of cycles of lengths greater than $2^n$ remains an open question. \\
\\
{\bf Aknowledgement.} This work was supported by the French Agency
 {\em Aire d\'eveloppement} through the project {\em Calcul Parall\`ele} and
 by the Project NTIC of the University of Ngaound\'er\'e.

\bibliographystyle{fplain}

\end{document}